\documentclass[twocolumn,english,aps,arxiv,superscriptaddress]{revtex4-1}
\usepackage{natbib}
\usepackage{grffile}
\usepackage{gensymb}
\usepackage[latin9]{inputenc}
\usepackage{color}
\usepackage{babel}
\usepackage{amsmath}
\usepackage{amsfonts}
\usepackage{graphicx}
\usepackage{epstopdf}
\usepackage{multirow}
\usepackage{cancel}
\PassOptionsToPackage{normalem}{ulem}
\usepackage{ulem}
\usepackage[unicode=true, bookmarks=false, breaklinks=false,pdfborder={0 0 1},colorlinks=false] {hyperref}
\usepackage{breakurl}

\begin{document}

\title{Single crystal growth, chemical defects, magnetic and transport properties of antiferromagnetic topological insulators (Ge$_{1-\delta-x}$Mn$_x$)$_2$Bi$_2$Te$_5$ ($x\leq 0.47$, $0.11 \leq \delta \leq 0.20$)}

\author{Tiema Qian}
\affiliation{Department of Physics and Astronomy and California NanoSystems Institute, University of California, Los Angeles,
CA 90095, USA}

\author{Chaowei Hu}
\affiliation{Department of Physics and Astronomy and California NanoSystems Institute, University of California, Los Angeles,
CA 90095, USA}

\author{Jazmine C. Green}
\affiliation{Department of Physics and Astronomy and California NanoSystems Institute, University of California, Los Angeles,
CA 90095, USA}

\author{Erxi Feng}
\affiliation{Neutron Scattering Division, Oak Ridge National Laboratory, Oak Ridge,
Tennessee 37831, USA}

\author{Huibo Cao}
\affiliation{Neutron Scattering Division, Oak Ridge National Laboratory, Oak Ridge,
Tennessee 37831, USA}

\author{Ni Ni}
\email{Corresponding author: nini@physics.ucla.edu}
\affiliation{Department of Physics and Astronomy and California NanoSystems Institute, University of California, Los Angeles, CA 90095, USA}

\begin{abstract}

Magnetic topological insulators provide a platform for emergent phenomena arising from the interplay between magnetism and band topology. Here we report the single crystal growth, crystal structure, magnetic and transport properties, as well as the neutron scattering studies of topological insulator series (Ge$_{1-\delta-x}$Mn$_x$)$_2$Bi$_2$Te$_5$ ($x\leq 0.47$, $0.11 \leq \delta \leq 0.20$). Upon doping up to $x = 0.47$, the lattice parameter $c$ decreases by 0.8\%, while the lattice parameter $a$ remains nearly unchanged. Significant Ge vacancies and Ge/Bi site mixing are revealed via elemental analysis as well as refinements of the neutron and X-ray diffraction data, resulting in holes dominating the charge transport. At $x = 0.47$, below 10.8 K, a bilayer A-type antiferromagnetic ordered state emerges, featuring an ordered moment of 3.0(3) $\mu_{B}$/Mn at 5 K, with the $c$ axis as the easy axis. Magnetization data unveil a much stronger interlayer antiferromagnetic exchange interaction and a much smaller uniaxial anisotropy compared to MnBi$_{2}$Te$_{4}$. We attribute the former to the shorter superexchange path and the latter to the smaller ligand-field splitting in (Ge$_{1-\delta-x}$Mn$_x$)$_2$Bi$_2$Te$_5$. Our study demonstrates that this series of materials holds promise for the investigation of the Layer Hall effect and quantum metric nonlinear Hall effect.

\end{abstract}
\date{\today}
\maketitle
\section{Introduction}

The discovery of magnetic topological insulators (TIs) marks an important breakthrough in condensed matter physics in the past decade. When magnetism is introduced in TIs and breaks the time-reversal symmetry that protects the gapless Dirac surface states, a gapped surface state and dissipationless quantized edge conduction may appear. Therefore magnetic TIs can host a set of emergent phenomena such as quantum anomalous Hall effect, Axion insulating state and quantum magnetoelectric effect\cite{tokura2019magnetic,he2018topological,liu2016quantum,wang2015quantized}. Among the magnetic TIs, MnBi$_{2n}$Te$_{3n+1}$ family with alternating [MnBi$_2$Te$_4$] septuple layer (SL), and $(n-1)$[Bi$_2$Te$_3$] quintuple layer (QL) is the first family that hosts intrinsic magnetism rather than introduced by doping \cite{lee2013crystal,rienks2019large,zhang2019topological,li2019intrinsic, otrokov2019prediction,gong2019experimental,lee2019spin,yan2019crystal,zeugner2019chemical,otrokov2019unique, aliev2019novel,147,ding2020crystal,wu2019natural,shi2019magnetic,tian2019magnetic,yan2020type,gordon2019strongly,deng2020high-temperature}. MnBi$_{2n}$Te$_{3n+1}$ goes from an A-type antiferromagnetic (AFM) state ($n\le 3$) to a ferromagnetic state ($n \ge 4$), with magnetic moment pointing out-of-plane. The van der Waals nature makes it easy to exfoliate a bulk crystal into a thin-film device, in which quantized anomalous Hall conductance \cite{liu2020robust,deng2020quantum} and electric-field tuned Layer Hall effect \cite{gao2021layer} are experimentally achieved in odd-layer and even-layer MnBi$_2$Te$_4$ devices, respectively. 

The discovery of MnBi$_{2n}$Te$_{3n+1}$ was inspired by the existence of the nonmagnetic XBi$_{2n}$Te$_{3n+1}$ (X = Ge, Sn, Pb) series which have already been synthesized for decades \cite{shelimova2004crystal}. XBi$_{2n}$Te$_{3n+1}$ are previously known thermoelectric materials, and recently attracted research interest due to their non-trivial band topology \cite{neupane2012topological,okamoto2012observation,kuroda2012experimental}. When nonmagnetic X atoms are replaced by Mn, the quasi-metastable MnBi$_{2n}$Te$_{3n+1}$ compounds can be made in a very narrow temperature region \cite{1813}. To the XTe-rich end of the XTe-Bi$_2$Te$_3$ phase diagram, besides XBi$_{2n}$Te$_{3n+1}$, thicker layered structures with more X in one building block exist. For example, X$_2$Bi$_2$Te$_5$ (X = Ge, Sn, Pb), abbreviated as the 225 phase, is made of nonuple layers (NL) while X$_3$Bi$_2$Te$_6$ (X = Ge, Sn, Pb) phase consists of undecuple layers\cite{shelimova2004crystal,kuropatwa2012thermoelectric,chatterjee2015solution,matsunaga2007structures}. The NL of X$_2$Bi$_2$Te$_5$ can be seen as inserting an additional XTe layer into XBi$_2$Te$_4$ SL, as shown in Fig. 1(a). Given the close structural correspondence between MnBi$_2$Te$_4$ and XBi$_2$Te$_4$, one may suspect Mn$_2$Bi$_2$Te$_5$ (Mn225) and Mn$_3$Bi$_2$Te$_6$ to exist, being potential candidates of intrinsic magnetic topological insulators. Indeed theoretical calculation has indicated Mn225 to be an intrinsic magnetic topological insulator that could host dynamical axion field \cite{li2020intrinsic, zhang2020large, eremeev2022magnetic, li2023stacking, tang2023intrinsic}. However, the successful growth of pure Mn225 phase is very challenging, hindering the investigation of its intrinsic physical properties \cite{yan2022vapor, cao2021growth}. For example, only a few layers of Mn225 phase were found embedded inside the MnBi$_2$Te$_4$ pieces in chemical vapor transport (CVT) growths while the Mn225 single crystals obtained via the self-flux growth might show significant contamination from the MnBi$_2$Te$_4$ phase. 

In this paper, we report the growth, crystal and magnetic structures, as well as the transport and thermodynamic properties of high quality (Ge$_{1-\delta-x}$Mn$_x$)$_2$Bi$_2$Te$_5$ ($x\leq 0.47$, $0.11 \leq \delta \leq 0.20$) single crystals. While our attempt to grow pure Mn225 single crystals is not successful using both the CVT and flux growth methods, pure (Ge$_{1-\delta}$)$_2$Bi$_2$Te$_5$ (Ge225) single crystals were made by the flux method using Te as the self flux while (Ge$_{1-\delta-x}$Mn$_x$)$_2$Bi$_2$Te$_5$ (GeMn225) with $x\leq 0.47$ can be grown by the CVT method \cite{CVT124,yan2022vapor}. The wavelength-dispersive X-
ray spectroscopy (WDS) measurements as well as the refinements of the powder X-ray diffraction (PXRD) and single-crystal neutron diffraction data indicate the presence of significant Ge vacancies of $0.11 \leq \delta \leq 0.20$, leading to holes dominating the electrical transport. We find that GeMn225 shows a $T_N = 10.8$ K at $x=0.47$ with a spin flop transition at 2.0 T when $H\parallel c$. Our neutron analysis of the $x=0.47$ compound suggests negligible amount of Mn$_{\rm{Bi}}$ antisite formation and a bilayer A-type AFM with a refined Mn moment of 3.0(3) $\mu_B$ at 5 K.

\section{Experimental methods}

Ge225 single crystals were grown using the self-flux method with Te as the flux. Ge chunks, Bi chunks and Te chunks were mixed at the ratio of Ge : Bi : Te = 2 : 2 : 8 in an alumina crucible and sealed in an evacuated quartz tube. The ampule was first heated to 1000 $^\circ$C overnight, then quickly cooled to 600 $^\circ$C before it was slowly cooled to 520 $^\circ$C in 3 days. At last, single crystals were separated from the flux by a centrifuge. Large and shiny mm-sized single crystals were obtained using this method.
\begin{figure}
    \centering
    \includegraphics[width=3.5in]{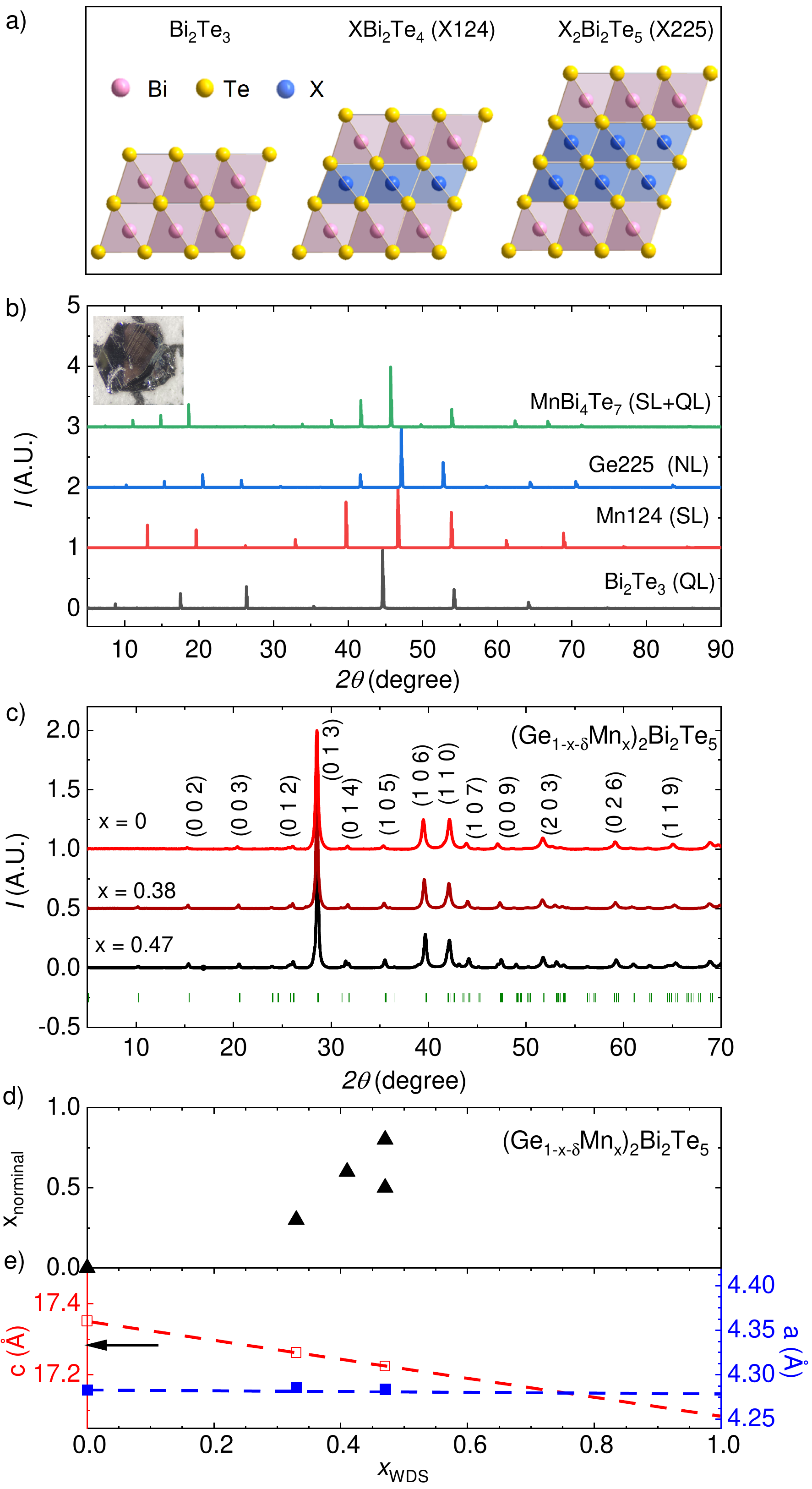}
    \caption{(a) Crystal structures of Bi$_2$Te$_3$ QL, XBi$_2$Te$_4$ SL and X$_2$Bi$_2$Te$_5$ NL. (b) (0 0 $L$) Bragg peaks of different X-Bi-Te series. Inset: an as-grown hexagonal single crystal of GeMn225 ($x = 0.47$) against a mm-grid. (c) PXRD of (Ge$_{1-\delta-x}$Mn$_x$)$_2$Bi$_2$Te$_5$. (d) $x_{\rm{nominal}}$ vs. $x_{\rm{WDS}}$ when growing GeMn225. (e) Lattice parameters $a$ and $c$ of (Ge$_{1-\delta-x}$Mn$_x$)$_2$Bi$_2$Te$_5$, dashed line shows the linear fit of lattice parameters.}
    \label{XRD}
\end{figure}{}

Our flux-growth trials of the Mn-doping series using Te self flux did not yield the 225 phase. However, our CVT growth trials using MnI$_2$ as the transport agent resulted in high quality GeMn225 single crystals. Mn pieces, Ge chunks, Bi chunks, Te chunks and I$_2$ pieces were mixed at the ratio given in Table I, loaded and sealed in a quartz tube under vacuum. The tube was placed vertically in a box furnace and slowly heated to 1000 $^\circ$C overnight. It was then moved to a horizontal tube furnace where the low-temperature and high-temperature were set to be 520$^\circ$C and 540$^\circ$C, with the starting material on the high-temperature end. The cold-end temperature was selected as 520 $^{\circ}$C since it was the synthesis temperature reported for pure Ge225 in solid-state reaction \cite{kuropatwa2012thermoelectric}. 
After two weeks, GeMn225 single crystals were taken out and rinsed with distilled water to remove the iodide impurities.

\begin{table*}[]
\setlength{\tabcolsep}{5pt}
\caption{Summary of the (Ge$_{1-\delta-x}$Mn$_x$)$_2$Bi$_2$Te$_5$ series. All doped compounds are grown by the CVT method with MnI$_2$ as the transport agent while the parent compound is made by the flux method as discussed in the text. $^*$: the ratio of Ge$_{1-x}$Mn$_x$Te : Bi$_2$Te$_3$ : MnI$_2$. $a$ and $c$ are the lattice parameters. $T_N$ is the AFM transition temperature. $p_1$ is the charge carrier density calculated from Hall measurements via $p_1=B/e\rho_{yx}$, $p_2$ is the charge carrier density estimated by $p_2=2\delta/A$, where $A$ is the unit cell volume in cm$^3$.}
\label{synthesis}
\begin{tabular}{cccccccccc}
\hline
$x$$_{\rm{nominal}}$  &ratio$^*$& WDS& $x$& $\delta$ & $a$ (\AA)& $c$ (\AA) & $T_{N}$ (K) & $p_1$(cm$^ {-3}$) & $p_2$(cm$^ {-3}$) \\
\hline\hline
0 & (Flux) & Ge$_{1.59(3)}$Bi$_{1.94(2)}$Te$_5$ &0 &0.20(2)& 4.283(1)&  17.352(1) & NA & 9.2$\times10^{20}$&2.9$\times10^{21}$\\
0.3 & 2:1:1 & Mn$_{0.65(1)}$Ge$_{1.07(1)}$Bi$_{1.99(2)}$Te$_5$ &0.33(1) &0.14(1)& 4.286(1)& 17.263(1) & 6.0 &4.7$\times10^{20}$&1.7$\times10^{21}$\\
0.5 & 2:1:1 & Mn$_{0.94(1)}$Ge$_{0.82(3)}$Bi$_{2.02(2)}$Te$_5$ &0.47(1) &0.12(2)& 4.284(1)& 17.223(1) & 10.8 &1.6$\times10^{20}$&1.3$\times10^{21}$\\
0.6 & 3:1:1 & Mn$_{0.83(1)}$Ge$_{0.93(2)}$Bi$_{1.99(2)}$Te$_5$ &0.41(1) &0.12(1)& 4.284(1)& 17.225(1) & 10.0 &&\\
0.8 & 5:1:1 & Mn$_{0.94(3)}$Ge$_{0.85(1)}$Bi$_{1.99(2)}$Te$_5$ &0.47(1) &0.11(2)& 4.285(1)& 17.238(1) & 11.0 &&\\\hline

\end{tabular}
\end{table*}

To identify the pieces of the 225 phase, (0 0 $L$) reflections were collected on both the top and bottom surfaces of single crystals using a PANalytical Empyrean diffractometer equipped with Cu K$\alpha$ radiation. Following this, we performed PXRD for further impurity checking and structural refinement. Then WDS measurements were conducted to obtain the elemental analysis of the samples, specifically the Mn level $x$. Magnetization data were collected in a Quantum Design (QD) Magnetic Properties Measurement System (MPMS). Specific heat and electrical transport measurements were made inside a QD DynaCool Physical Properties Measurement System (PPMS). Electrical contacts were made to the sample using Dupont 4922N silver paste to attach Pt wires in a six-probe configuration. To eliminate unwanted contributions from mixed transport channels, electrical resistivity ($\rho_{xx}$) and Hall ($\rho_{yx}$) data were collected while sweeping the magnetic field from -9 T to 9 T. The data were then symmetrized to obtain $\rho_{xx}(H)$ using $\rho_{xx}(H)=(\rho_{xx}(H)+\rho_{xx}(-H))/2$ and antisymmetrized to get $\rho_{yx}(H)$ using $\rho_{yx}(H)=(\rho_{yx}(H)-\rho_{yx}(-H))/2$. The magnetoresistance is defined as MR $= (\rho_{xx}(H)-\rho_{xx}(0))/\rho_{xx}(0)$. In our measurement geometry, the positive slope of $\rho_{yx}(H)$ suggests hole carriers dominate the transport. Single-crystal neutron diffraction was performed for the $x$ = 0.47 sample at 5 K and 0 T on the HB-3A DEMAND single-crystal neutron diffractometer located at Oak Ridge National Laboratory\cite{chakoumakos2011four}. Both the neutron and X-ray diffraction data were refined using the Fullprof suit \cite{rodriguez1993recent}.

\section{Experimental Results} 

\subsection{Growth optimization and phase characterization}

Our CVT growth trials of the GeMn225 phase started with an elemental ratio such that XTe : Bi$_2$Te$_3$ : MnI$_2$ = $m:1:1$, where X = (Ge$_{1-x}$Mn$_x$). As we increased the Mn concentration in X, higher $m$ for extra XTe became necessary to yield the 225 phase. Our optimal trials that gave high quality GeMn225 single crystals are listed in Table \ref{synthesis}. Ge225 and GeMn225 crystals can grow up to a lateral size of several mm with a thickness about a hundred micron in two weeks. All crystals obtained from the CVT growth process exhibit a hexagonal-plate shape, with clearly defined edges indicating the as-grown $a$ and $b$ axes. In the inset of Fig. \ref{XRD} (b), an image of a (Mn$_{0.47}$Ge$_{0.41}$)$_2$Bi$_2$Te$_5$ single crystal against the mm-grid is shown .

The GeMn225 phase was first confirmed by checking the (0 0 $L$) reflections in the surface XRD patterns. Because the (0 0 $L$) spectrum depends solely on the periodic unit along $c$ axis, $i.e.$, the thickness of the NL layer, it can be well distinguished from that of [MnBi$_2$Te$_4$] SL, [Bi$_2$Te$_3$] QL or their combinations. A comparison of the (0 0 $L$) reflections of various materials is shown in Fig. \ref{XRD}(b), revealing the increasing thickness of the repeating layer(s) from QL, SL, NL to QL+SL. The PXRD patterns are shown and indexed in Fig. \ref{XRD}(c). No clear impurity phases were identified.

The Mn doping levels obtained via the WDS measurements are summarized in Table I. 
 These values suggest the highest doping level of Mn remains to be around $x=0.47$ in GeMn225 despite the nominal $x$ in the starting materials being much higher than 0.47. Based on the experience stated above, we also attempted pure Mn225 growth with extra MnTe. High-$m$ trials such as Mn : Bi : Te : I = 11 : 2 : 13 : 2 at various growth temperatures yield only MnBi$_2$Te$_4$ and/or Bi$_2$Te$_3$. Via both flux and CVT methods, we were unable to obtain pure Mn225 single crystals. So for this GeMn225 phase to appear stably in CVT growth, we conclude that there exists a substitution limit of Mn on Ge as indicated in Fig. \ref{XRD}(d).

The refined lattice parameters $a$ and $c$ are plotted in Fig. \ref{XRD}(e) against the $x$ values that are determined by WDS. The lattice parameter $a$ remains almost unchanged while the lattice parameter $c$ decreases by 0.8\% from $x=0$ to $x=0.47$. Assuming the Vegard's law, the extrapolation of the lattice parameters with $x$ allows us to predict the lattice parameters for pure Mn225. The data suggest Mn225 has $a = 4.27 \AA$ and $c = 17.1 \AA$, which is consistent with the previous report \cite{cao2021growth}.

\subsection{Magnetic and Transport properties of (Ge$_{1-\delta-x}$Mn$_x$)$_2$Bi$_2$Te$_5$ single crystals }

\begin{figure}[!t]
    \centering
    \includegraphics[width=3.4in]{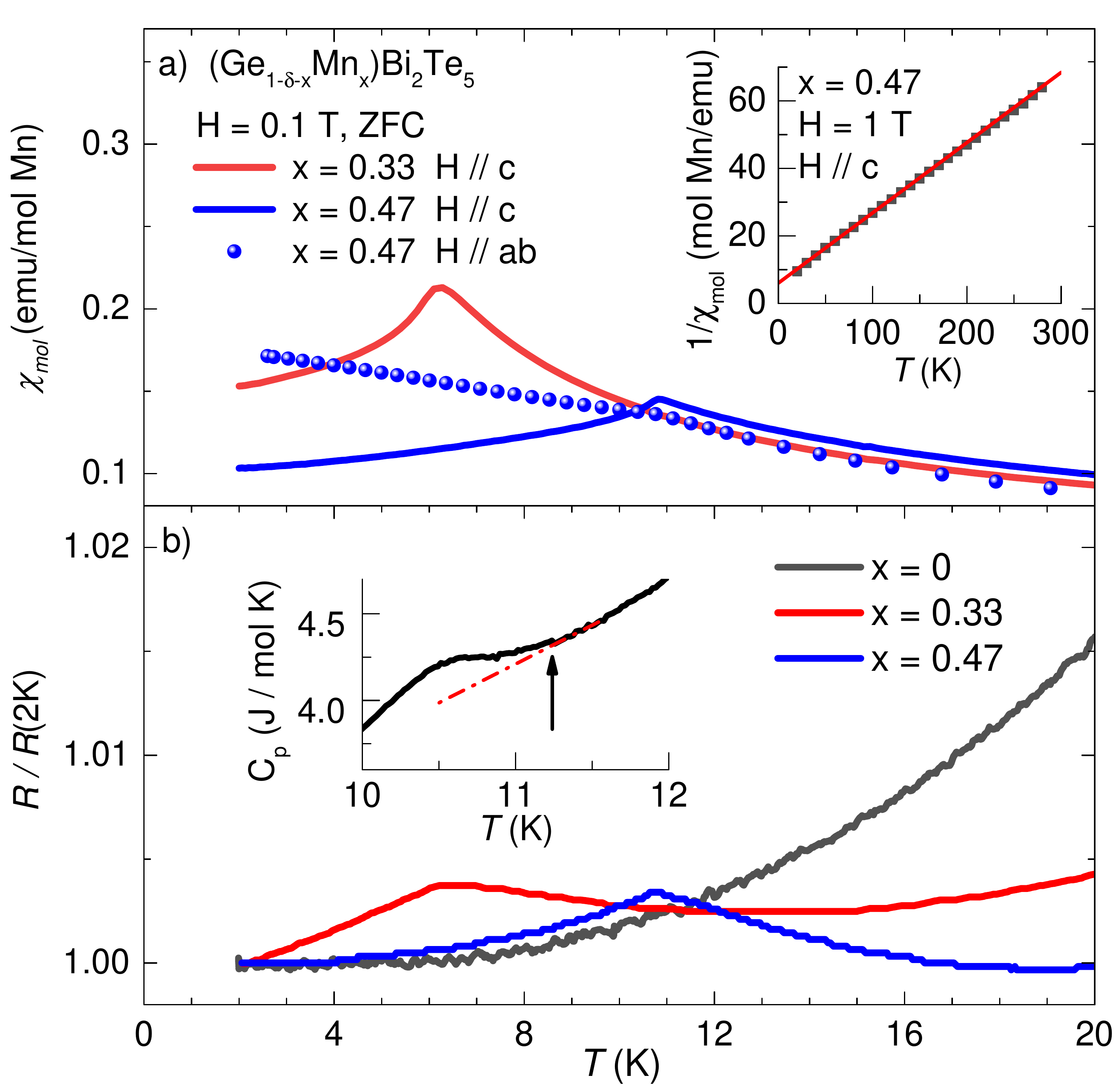}
    \caption{Thermodynamic and transport properties of GeMn225. (a) Temperature-dependent magnetic susceptibility under 0.1 T for different doping levels and direction. Inset: inverse magnetic susceptibility measured at 1 T above $T_N$. Curie-Weiss fit is shown in solid line. (b) Normalized temperature-dependent electrical resistivity with current along the $ab$ plane for different doping levels. Inset: temperature-dependence of specific heat of the $x=0.47$ sample with the criterion to determine $T_N$.}
    \label{MnGe-doping}
\end{figure}

To investigate the effect of Mn doping, we conducted thermodynamic and transport measurements. The Mn concentrations measured via WDS are utilized in the analysis of the magnetic and specific heat data and will be referenced throughout the paper. In Fig. \ref{MnGe-doping} (a), the temperature-dependent magnetic susceptibility, $\chi(T)$, measured at 0.1 T, reveals a kink feature at 6.0 K and 10.8 K for the $x = 0.33$ and $x = 0.47$ samples, respectively, indicating magnetic ordering at low temperatures. As the temperature decreases, $\chi(T)$ continues to rise below the ordering temperature for $H \parallel ab$, while it decreases for $H \parallel c$, indicating AFM ordering with the easy axis along the $c$ direction. The Curie-Weiss fit of the inverse susceptibility measured at 1 T (inset of Fig. \ref{MnGe-doping} (a)) yields a Curie temperature of -12 K that suggests strong in-plane ferromagnetic fluctuation and an effective moment of 6.0 $\mu_B$/Mn that is consistent with Mn$^{2+}$'s effective moment. Figure \ref{MnGe-doping} (b) presents the normalized temperature-dependent longitudinal resistivity with the current along the $ab$ plane, $\rho_{xx}(T)/\rho_{xx}$(2 K). While the resistivity in the undoped one exhibits a monotonic decrease upon cooling, the sharp drop in resistivity for the $x = 0.33$ and $0.47$ samples suggests suppressed spin scattering upon entering the ordered state, implying parallel in-plane spin alignment. The inset of Fig. \ref{MnGe-doping} (b) presents the specific heat data of the $x=0.47$ compound, revealing an anomaly associates with the AFM transition emerging at 10.8 K, in line with other measurements.

{}

\begin{figure}
    \centering
    \includegraphics[width=3.5in]{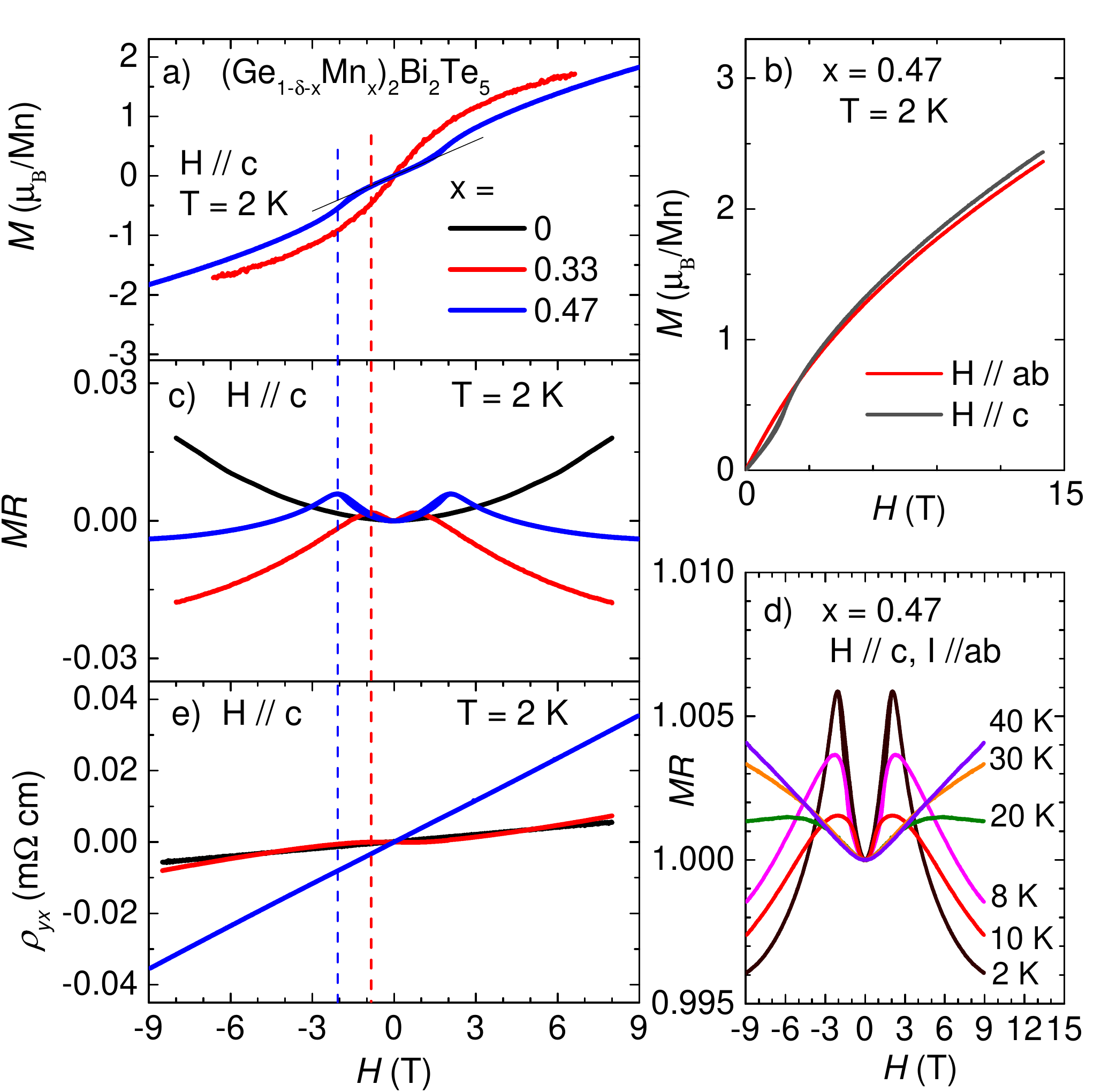}
    \caption{(a) Isothermal magnetization at 2 K of GeMn225 with $H \parallel c$. (b) Isothermal magnetization measured at 2 K up to 14 T for the $x = 0.47$ sample with $H \parallel ab$ and $H \parallel c$. (c) Field-dependence of MR of GeMn225. (d) MR at various temperatures for the $x = 0.47$ sample. (e) Field-dependence of Hall resistivity. Dotted lines refer to $H_{sf}$ in the $x=0.33$ and 0.47 samples.}
    \label{GeMn0.5}
\end{figure}{}

The evolution of magnetism under external fields and its coupling with charge carriers are presented in Fig. \ref{GeMn0.5}. Figure \ref{GeMn0.5} (a) shows their isothermal magnetization for $H\parallel c$. While both curves exhibit AFM behavior, a clear spin-flop transition feature appears in the $x = 0.47$ sample at about $H_{\rm{sf}}=2.0$ T. This value is lower than that of 3.3 T in MnBi$_2$Te$_4$, yet significantly higher than the 0.2 T observed in MnBi$_4$Te$_7$. No sign of spin-flop transition is observed for $H\parallel ab$ (Fig. \ref{GeMn0.5} (b)), indicating the $c$-axis as the easy axis. Magnetization in both doped samples is about 1.8 $\mu_B$/Mn at 7 T. For the $x = 0.47$ sample, $M$ reaches to 2.4 $\mu_B$/Mn at 14 T with no sign of saturation, as shown in Fig. \ref{GeMn0.5}(b). This value is less than half of the expected Mn moment of 5 $\mu_B$/Mn, suggesting that the saturation field is much higher than 14 T.

Figures \ref{GeMn0.5} (c) and (d) depict the MR data. The MR of the $x=0$ sample exhibits a parabolic field dependence while it peaks at $H_{\rm{sf}}=0.7$ T and $H_{\rm{sf}}=2.0$ T for the $x=0.33$ and $0.47$ compounds, respectively. Above $H_{\rm{sf}}$, the MR displays a negative slope as the spin disorder scattering gradually diminishes with increasing field. This negative slope in MR persists at elevated temperatures up to 30 K, as illustrated in Fig. \ref{GeMn0.5} (d), suggesting significant spin fluctuation above the ordering temperature in this doping series. Figure \ref{GeMn0.5} (e) presents the field-dependent Hall resistivity. Its positive slope with magnetic field suggests holes dominate the transport. The carrier concentrations are in the order of 10$^{20}$ cm$^{-3}$ and decrease with higher Mn doping, as summarized in Table I. This is in sharp contrast with the previous report on Mn225 where electrons dominate the transport \cite{cao2021growth}.

\subsection{Crystal and Magnetic Structure}

If free of defects, the stoichiometry of Ge : Bi : Te would be 2 : 2 : 5 for Ge225. However, as indicated in Table I, WDS measurements reveal a deficiency of Ge, with only 1.59 Ge atoms present in Ge225. Meanwhile, a (Ge+Mn) deficiency in Mn-doped samples also exists, where (Ge+Mn) $\sim$ 1.7. In order to better understand the crystal and magnetic structure of this family, particularly regrading the types of defects present and the specific sites where Mn is doped, we have performed both single crystal neutron diffraction for the $x = 0.47$ sample and PXRD for the Ge225 sample. The $2d$ and $2c$ sites are where (Mn/Ge/Bi) cations can reside, forming four cation layers. In each NL, atoms on the $2d$ site make the inner two cation layers, while those on the $2c$ site constitute the outer two cation layers.

\subsubsection{Magnetic structure revealed through Neutron Diffraction Analysis}
Since neutron diffraction is quite sensitive to Mn atoms in the Mn-Bi-Te systems due to the negative scattering length of Mn\cite{ding2020crystal, ding2021neutron}, we first measure the $x = 0.47$ crystal using neutron diffraction to determine the magnetic structure and whether Mn is doped onto the $2d$ or $2c$ site. 

No additional Bragg peaks are observed below $T_N$, indicating a magnetic propagation vector of (0 0 0). The right inset of Fig. 4 shows the intensity of the (0 1 0) and (0 0 4) peaks below and above $T_N$. Upon entering the ordered state, the (0 1 0) peak increases, indicating formation of spin order perpendicular to $b$ axis. An unchanged intensity in (0 0 4) peak, on the other hand, indicates likely no spin component perpendicular to $c$ axis. This points to an easy axis along $c$ axis without spin tilting, consistent with our magnetic property measurements. So given its AFM nature and the crystal space group $P\overline{3}m1$ (No. 164), the highest magnetic symmetry $P\overline{3}'m'1$ with the ordered moment along $c$ can be concluded and used to fit the collected neutron data. Because Ge and Bi have similar scattering lengths for neutrons, it is difficult to differentiate Ge and Bi on the same site. For simplicity, in our refinement, we assume Ge and Bi each occupy either $2d$ or $2c$ site, with our primary focus being on Mn distributions. Note in reality Ge and Bi mixing is expected, which we will discuss later through X-ray diffraction analysis. We examine three possible scenarios, stacking A with Mn on the $2d$ site, stacking B with Mn on the $2c$ site, and a mixed stacking where Mn can go into either site (Table S1 - S3) \cite{supplement}. In all three scenarios, we refine the occupancy of Ge and Mn, as well as the moment of Mn. Our refinement demonstrates that the scenario where all Mn atoms reside on the $2d$ site yields the highest goodness-of-fit value. Therefore, within the resolution of our measurement, we conclude that Mn is doped onto the $2d$ site, with Mn residing on the inner two layers, as shown in the left inset of Fig. 4. Our refinement indicates parallel alignment of spins within the $ab$ plane, with spins in adjacent layers being antiparallel to each other \cite{supplement}.  

\begin{table}[]
\setlength{\tabcolsep}{3pt}
\label{table:XrayGe225}
\renewcommand{\arraystretch}{1.5}
\caption{Refined crystal structural parameters for the parent compound Ge225 based on the PXRD data measured at 300 K. The refinement is constrained by the WDS result. Number of reflections: 6474; $R_F = 8.42\%$; $\chi^2 = 46.4$.}
\centering
\begin{tabular}{ccccccc}
\hline
Atom & site  & $x$     & $y$     & $z$          & occ.    \\
\hline\hline
Ge1  & $2d$ & 1/3           &2/3     & 0.1043(3)     & 0.640(6)  \\
Bi1  & $2d$ & 1/3          &2/3     & 0.1043(3)     & 0.361(6)  \\
Ge2  & $2c$ & 0                &0          & 0.3260(2)     & 0.161(6)  \\   
Bi2  & $2c$ & 0                &0          & 0.3260(2)     & 0.639(6)  \\
Te1  & $1a$ & 0                &0          & 0             & 1         \\
Te2  & $2d$ & 1/3           &2/3     & 0.2037(3)     & 1         \\
Te3  & $2d$ & 1/3           &2/3     & 0.4243(2)     & 1         \\\hline\hline
\end{tabular}
\end{table}

\begin{table}[]
\setlength{\tabcolsep}{3pt}
\label{table:Neutron}
\renewcommand{\arraystretch}{1.5}
\caption{Refined magnetic and crystal structural parameters for the $x=$0.47 sample based on the single crystal neutron diffraction data measured at 5 K. The refinement is constrained by the WDS result. Number of reflections: 38; $R_F = 12.8\%$; $\chi^2 = 7.12$.}
\centering
\begin{tabular}{ccccccc}
\hline
Atom & site & $x$     & $y$     & $z$          & occ. & Moment at 5 K             \\
\hline\hline
Ge1  & $2d$ &1/3  & 2/3  & 0.099(6)   & 0.17       \\
Mn1  & $2d$ &1/3  & 2/3  & 0.099(6)   & 0.47   & 3.0(3) $\mu_{B}$/Mn            \\
Bi1  & $2d$ &1/3  & 2/3  & 0.099(6)   & 0.36       \\
Ge2  & $2c$ & 0      & 0       & 0.316(2)   & 0.23          \\
Bi2  & $2c$ & 0      & 0       & 0.316(2)   & 0.64          \\
Te1  & $1a$ & 0      & 0       & 0          & 1          \\
Te2  & $2d$ & 1/3 & 2/3  & 0.792(3)   & 1          \\
Te3  & $2d$ & 1/3 & 2/3  & 0.426(3)   & 1          \\
\hline\hline
\end{tabular}
\end{table}

\subsubsection{Vacancies and Bi/Ge site mixing in Ge225}

According to the WDS measurements, Ge225 samples may exhibit vacancies. If we assume the presence of vacancies and Bi/Ge site mixing, we can write down: 
\begin{align} 
f_{2d}=\mbox{Ge}^{occ}_{2d} f_{\rm{Ge}} + \mbox{Bi}^{occ}_{2d} f_{\rm{Bi}} + V_{2d} \times 0 \\
~f_{2c}=\mbox{Ge}^{occ}_{2c} f_{\rm{Ge}} + \mbox{Bi}^{occ}_{2c} f_{\rm{Bi}} + V_{2c} \times 0.  
\end{align}
Here, $f$ is the atomic scattering factor, $^{occ}$ refers to the element occupancy, $V$ is the amount of vacancy.

Two extreme structural models are used to obtain $f_{2d}$ and $f_{2c}$. In model 1, Ge occupies both $2d$ and $2c$ sites while in model 2, Bi occupies both. The refinements show that in model 1, $\mbox{Ge}_{2d}^{occ1}$ and $\mbox{Ge}_{2c}^{occ1}$ equals 1.59 and 1.79, and in model 2, $\mbox{Bi}_{2d}^{occ2}$ and $\mbox{Bi}_{2c}^{occ2}$ equals 0.58 and 0.66 \cite{supplement}. Since regardless of the occupancy model employed, the scattering cross section of an individual site should be the same, we can write down:
\begin{align} 
\mbox{Site} ~2d: f_{2d}=1.59 f_{\rm{Ge}} = 0.58 f_{\rm{Bi}}\\
\mbox{Site} ~2c: f_{2c}=1.79 f_{\rm{Ge}} = 0.66 f_{\rm{Bi}},
\end{align}
which lead to $f_{\rm{Bi}}/f_{\rm{Ge}} = 2.7$. This number is close to the atomic number ratio between Bi and Ge, 2.6. By plugging this ratio into Eq. (1), we obtain :
\begin{align}
\mbox{Ge}^{occ}_{2d}  + 2.7\mbox{Bi}^{occ}_{2d}  = 1.59 
\end{align}
By plugging the ratio into Eq. (2) and with $\mbox{Ge}^{occ}_{2c}=1-V_{\rm{Ge}}-\mbox{Ge}^{occ}_{2d}$ and $\mbox{Bi}^{occ}_{2c}=1-V_{\rm{Bi}}-\mbox{Bi}^{occ}_{2d}$ where $V_{\rm{Ge}}$ and $V_{\rm{Bi}}$ refer to the amount of vacancies for Ge or Bi, we get: 
\begin{align}
(1 - V_{\rm{Ge}} - \mbox{Ge}^{occ}_{2d}) + 2.7(1 - V_{\rm{Bi}} - \mbox{Bi}^{occ}_{2d}) = 1.79,
\end{align}
From Eqs. (5) and (6), we get :
\begin{align}
V_{\rm{Ge}} + 2.7 V_{\rm{Bi}} = 0.32
\end{align}
Therefore, PXRD also suggests vacancies in the compound. WDS measurements show $V_{\rm{Ge}}$ to be 0.20(2) and $V_{\rm{Bi}}$ is 0.03(1), which is consistent with Eq. (7).

Utilizing the aforementioned constraint and with the total amount of Ge and Bi set to their WDS values, the occupancy of Ge and Bi on each site is refined. The refinement returns the same goodnss-of-fit when assuming all vacancies on the $2d$ site (Table S6), $2c$ site (Table II), or distributed on both $2d$ and $2c$ sites (Table S7) \cite{supplement}. Based on the refinement of our neutron diffraction data, which suggests that Mn is doped on the $2d$ site, and considering the WDS measurements indicating that Mn atoms solely substitute Ge atoms, it is reasonable to infer that most Ge atoms occupy the $2d$ site. Our refinement shows that Ge atoms predominantly occupy site $2d$ when vacancies concentrate on site $2c$. The refined crystal structure is thus finalized in Table II. 

\subsubsection{Neutron Refinement of the $x = 0.47$ Sample}
With a better understanding of the crystal structure of Ge225, we turn back to the neutron diffraction data to work out the crystal and magnetic structure for the doped sample. Now we force the WDS values of Mn with Mn only replacing Ge on site $2d$, and set the distribution of Bi on both sites identical to that of the parent compound with all vacancies concentrating on site $2c$. The refined structure is shown in Table III. The ordered Mn moment at 5 K is refined to be 3.0(3) $\mu_B$.

\begin{figure}
    \centering
    \includegraphics[width=3.5in]{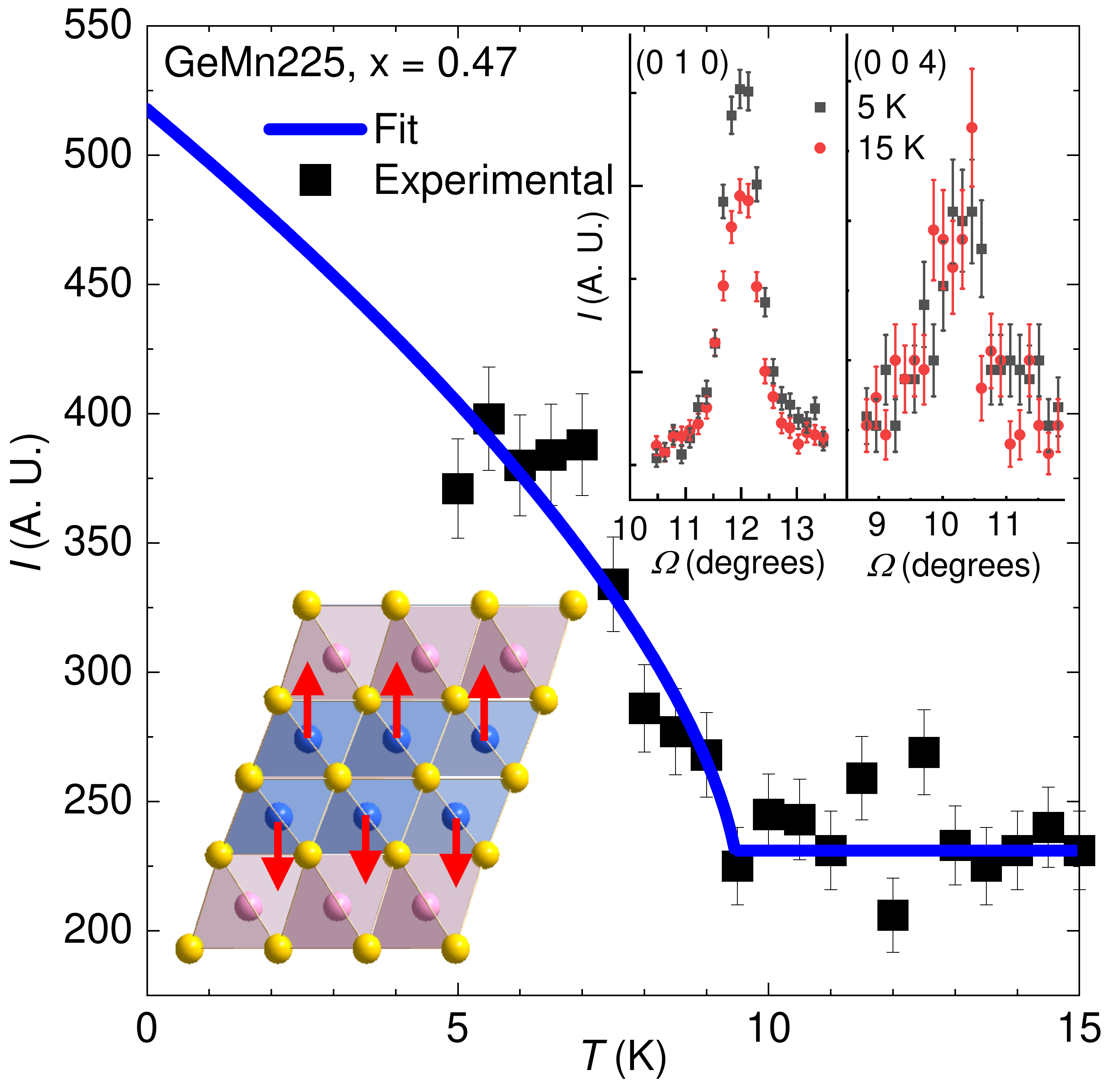}
    \caption{Magnetic order parameter at magnetic reflection (0 1 1) for the $x = 0.47$ sample measured at DEMAND. The blue line depicts the best fit using the mean-field power law, see text. Left inset: the magnetic structure. Right inset: the intensity of the (0 1 0) and (0 0 4) peak above and below $T_N$.}
    \label{order parameter}
\end{figure}

Figure \ref{order parameter} shows the magnetic order parameter, measured on the (0 1 1) refection up to 15 K for (Mn$_{0.47}$Ge$_{0.41}$)$_2$Bi$_2$Te$_5$. The solid line represents the fit to the mean-field power-law, 
\begin{equation}
I = A \left(\frac{T_{N} - T}{T_{N}}\right)^{2\beta} + B
\end{equation}
where $A$ is a constant, $B$ is the background and $\beta$ is the order parameter critical exponent. The best fit yields a Ne\'el temperature of $T_{N}$ = 9.5 K and a critical exponent of $\beta$ = 0.32(7), which is similar to that of MnBi$_2$Te$_4$ \cite{ding2020crystal}. Based on the fitting, we estimate the ordered moment at 0 K to be 4.5(7) $\mu_B$ per Mn, close to the expected value for Mn$^{2+}$. 

{}

\begin{figure*}[]
    \centering
    \includegraphics[width=7in]{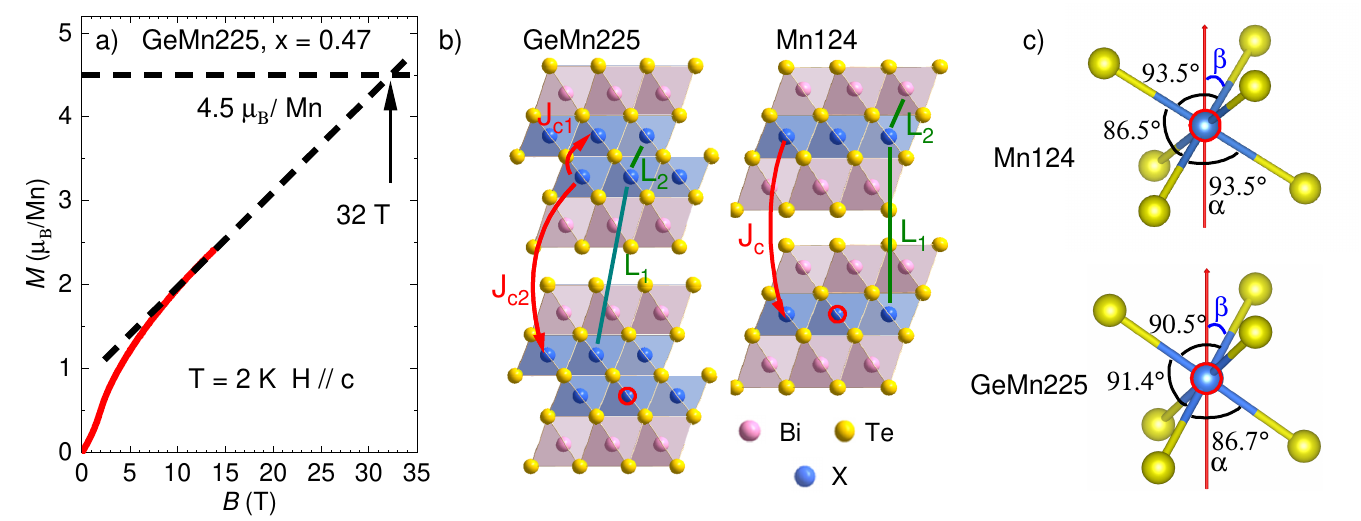}
    \caption{(a) Isothermal magnetization of the $x=0.47$ sample at 2 K. The red line is the experimental curve and the black dashed line is a linear extrapolation of the magnetization at 2 K, it reaches to 4.5 $\mu_{B}$/Mn at around 30 T. (b) Crystal structure of X225 (X = GeMn) and X124 (X = Mn). The superexchanges are indicated with red arrows, the nearest-neighbor Mn-Mn distances are highlighted with olive lines. (c) Distorted MnTe$_6$ octahedron, the building blocks of the magnetic layer in the 124 and 225 phases. The Mn atom in the center refers to the circled one in (b). Different Te-Mn-Te bond angels are shown, the asymmetry in GeMn225 arises from its asymmetric next-nearest-neighbor environment.}
    \label{dis}
\end{figure*}{}

\begin{table*}[]
\setlength{\tabcolsep}{2.5pt}
\label{table:Interactions}
\renewcommand{\arraystretch}{1}
\caption{Comparison in Mn-Bi-Te family. $SJ_c$ is the interlayer exchange coupling per Mn and $SK$ is the uniaxial magnetic anisotropy. $L_1$ and $L_2$ refer to the nearest-neighbor Mn-Mn interlayer distances shown in Fig. \ref{dis} (b). $\alpha$ and $\beta$ are the bond angles of the distorted MnTe$_6$ octahedron shown in the inset of Fig \ref{dis} (c).}
\centering
\begin{tabular}{ccccccc}
\hline
 Component & $SJ_{c1}$ (meV)    & $SK$ (meV)     & $L_1$ (\AA) & $L_2$ (\AA)  & $\alpha$ (\degree)& $\beta$ (\degree)\\
 \hline\hline
(Mn$_{0.47}$Ge$_{0.41}$)$_2$Bi$_2$Te$_5$    & 1.8   & 0.008    & 13.96 &  4.39 & 86.7/90.5& 52.4/55.1\\
(Mn$_{0.6}$Pb$_{0.4}$)Bi$_2$Te$_4$\cite{qian2022magnetic}&0.24   & 0.03   & 13.93 & 4.53 & 93.5 &  57.3      \\
MnBi$_2$Te$_4$ \cite{lai2021defect}         &0.26   & 0.09    & 13.64 & 4.51 & 93.5& 57.3\\
MnBi$_4$Te$_7$ \cite{hu2024recent}         &0.03   & 0.10    & 23.71 & & 93.5& 57.3\\
MnBi$_6$Te$_{10}$ \cite{hu2024recent}         &0.01   & 0.10    & 34.00 & & 93.2& 57.0\\
\hline\hline
\end{tabular}
\end{table*}

\section{Discussion}

 The presence of vacancies has profound impact in the transport properties of the 225 compounds. Research on Mn-Bi-Te systems indicates that defect-free compounds are charge-neutral, with carriers in actual samples being contributed by various defects \cite{hu2021growth}. Mainly, electron carriers are contributed by Bi$_{\rm{Mn/Ge}}$ and Te vacancies, whereas hole carriers are contributed by (Mn/Ge)$_{\rm{Bi}}$ and cation vacancies. This can be seen in the following defect chemistry for native Ge225:
\begin{align}
\rm{Ge_2Bi_2Te_5} \rightleftharpoons \rm{Ge}_{\rm{Bi}}^{\prime} + \textit{h}^{\bullet} + \rm{Bi}_{\rm{Ge}}^{\bullet} + \textit{e}^{\prime}
\end{align}
Indeed one Ge$_{\rm{Bi}}$ produces one hole while one Bi$_{\rm{Ge}}$ creates one electron. In the presence of Ge vacancies, we can write 
\begin{align}
\rm{(Ge_{1-\delta})_2Bi_2Te_4} \rightleftharpoons \delta\rm{V}_{\rm{Ge}}^{\prime\prime} + 2\delta\textit{h}^{\bullet}
\end{align}
Therefore, the hole carrier density can be estimated by calculating $2\delta/A$, where $A$ represents the unit cell volume with cm$^3$ as the unit. The carrier densities calculated through this defect analysis are denoted as $p_2$ and are summarized in Table I. As observed, the correspondence between $p_1$ and $p_2$ is reasonably good, especially for the $x=0$ and 0.33 samples. 

We may tentatively estimate the saturation field of the $x=0.47$ sample by assuming linear field dependence of $M$ above 14 T. The interpolation is shown in Fig. \ref{dis} (a). When the magnetization at 2 K reaches 4.5 $\mu_{B}$/Mn, the saturation field is estimated to be around 30 T. For a uniaxial antiferromagnet, long-range order requires either interlayer coupling, or uniaxial magnetic anisotropy. Due to the bilayer nature of 225, there exist two interlayer exchange couplings, as depicted in Fig. \ref{dis}(b): one is $J_{c1}$, representing the interlayer AFM coupling within each NL per Mn, the other is $J_{c2}$, denoting the interlayer coupling between adjacent NL per Mn. We can write down the full Hamiltonian in the ordered state, per Mn, as\cite{qian2022magnetic}:
\begin{equation}
\label{eq:9}
\begin{aligned}
E &=E_{0}+\frac{1}{2}x^2J_{c1}\mathbf{S}_{i}\cdot\mathbf{S}_{i+1} + \frac{1}{2}x^2J_{c2}\mathbf{S}_{i}\cdot\mathbf{S}_{i-1} \\
&-xKS_{z}^{2}-xg\mu_{B}\mathbf{S}_{i}\cdot\mathbf{H},
\end{aligned}
\end{equation}
where $g$ is the Lande factor, $\mathbf{S}_{i}$ represents the Mn spin under investigation, $\mathbf{S}_{i+1}$ is the Mn in the same NL as $\mathbf{S}_{i}$ while $\mathbf{S}_{i-1}$ is the Mn in the adjacent NL, $K$ is the magnetic anisotropy parameter per Mn and $S=5/2$. Since $\mathbf{S}_{i+1}$ and $\mathbf{S}_{i-1}$ represent identical spin, we can combine two exchange coupling as $J_c = 1/2(J_{c1}+J_{c2})$. The relationship between $J_c$ and $K$ is then,
\begin{align}
SK&=(g\mu_B/2)(H_{sf}^2/H_s^{\parallel c})\\
SJ_{c} &=(g\mu_B/4x)\left(H_s^{\parallel c}+H_{sf}^2\right/H_s^{\parallel c}),
\end{align}
Where $H_{sf}$ and $H_{s}$ is the spin flop field and saturation field. Using critical fields obtained above, we can get $SK$ = 8.0 $\mu$eV and $SJ_c$ = 1.8 meV. Table IV summarizes $SJ_c$, $K$, Mn-Mn distances and bond angles in Mn-Bi-Te family for comparison. Since $J_{c1} \gg J_{c2}$ owing to the much shorter superexchange path of $J_{c1}$ compared to $J_{c2}$, $J_{c1}$ can be approximated as $J_{c}$. $J_{c1}$ of (Mn$_{0.47}$Ge$_{0.41}$)$_2$Bi$_2$Te$_5$ is much larger than that of the MnBi$_{2n}$Te$_{3n+1}$ series. This is reasonable, given the much shorter Mn-Mn nearest-neighbor interlayer distance of 4.39 $\rm{\AA}$ in GeMn225 ($L_2$) compared to 13.86 $\rm{\AA}$ in MnBi${_2}$Te${_4}$ ($L_1$) or other Mn-Bi-Te compounds. Meanwhile, owing to a similar exchange path, $J_{c1}$ should be comparable to the coupling between the primary Mn site and the Mn$_{\rm{Bi}}$ antisite in MnBi$_2$Te$_4$. Indeed, the latter is responsible for the high full saturation field in MnBi$_2$Te$_4$ \cite{lai2021defect}. A much smaller magnetic anisotropy is obtained for (Mn$_{0.47}$Ge$_{0.41}$)$_2$Bi$_2$Te$_5$, compared to (Mn$_{0.6}$Pb$_{0.4}$)Bi$_2$Te$_4$, despite both have similar Mn occupancy and ordering temperature. This can be understood qualitatively by the bond angle analysis. As depicted in Table IV, both the Te-Mn-Te ($\alpha$) and Te-Mn-$z$ ($\beta$) angles exhibit a significant decrease in (Mn$_{0.47}$Ge$_{0.41}$)$_2$Bi$_2$Te$_5$ compared to MnBi$_{2n}$Te$_{3n+1}$. When the bond angles decrease, the ligand-field splitting will also become smaller due to a less overlap of wavefunctions, leading to a reduced magnetic anisotropy \cite{huisman1971trigonal, yan2021role}. This also explains why $SK$ remains similar across the MnBi$_{2n}$Te$_{3n+1}$ series (refer to Table IV), as the lattice environment of Mn remains consistent in these compounds.

When GeMn225 is exfoliated into even-NL or odd-NL thin flakes, both the inversion symmetry $\mathcal{P}$ and time reversal symmetry $\mathcal{T}$ are broken, while the combined $\mathcal{PT}$ symmetry is preserved. This symmetry condition is the same as the even-layer MnBi$_2$Te$_4$ device, where the Layer Hall effect \cite{gao2021layer} and quantum metric nonlinear Hall effect \cite{gao2023quantum} are discovered. Therefore, the bilayer A-type AFM and the non-trivial band topology nature of GeMn225 \cite{li2023stacking, zhang2020large} make it an excellent system for probing these emergent phenomena, eliminating the need to differentiate between even-NL or odd-NL devices.

\section{Conclusion} 

In summary, we have grown high-quality single crystals of (Ge$_{1-\delta-x}$Mn$_x$)$_2$Bi$_2$Te$_5$ with the doping level $x$ up to 0.47. Elemental analysis and diffraction techniques not only suggest Ge/Bi mixing, but also reveal the presence of significant Ge vacancies of $0.11 \leq \delta \leq 0.20$, being responsible for the holes dominating the charge transport. 
As $x$ increases, long-range AFM order with the easy axis along $c$ emerges at 6.0 K for the $x=0.33$ sample and at 10.8 K for the $x=0.47$ sample. Spin-flop transitions observed at 0.7 T for $x=0.33$ and 2.0 T for $x=0.47$. Our refinement of the neutron diffraction data of the $x=0.47$ sample suggests a bilayer A-type AFM structure with the ordered moment of 3.0(3) $\mu_B$/Mn at 5 K. Our analysis of the magnetization data reveals a much stronger interlayer AFM exchange interaction and a much reduced uniaxial magnetic anisotropy when contrasted with MnBi$_{2}$Te$_{4}$. We argue the former arises from the shorter superexchange path and the latter to be linked to the smaller ligand-field splitting in (Ge$_{1-\delta-x}$Mn$_x$)$_2$Bi$_2$Te$_5$. 
Our study illustrates that this series of materials always exhibit broken $\mathcal{P}$ and broken $\mathcal{T}$ symmetries yet preserved $\mathcal{PT}$ symmetry upon exfoliation into thin flakes, providing a platform to explore the Layer Hall effect and quantum metric nonlinear Hall effect.

\section*{Acknowledgments}
 We thank Randy Dumas at Quantum Design for high field magnetization measurements. Work at UCLA was supported by the U.S. Department of Energy (DOE), Office of Science, Office of Basic Energy Sciences under Award Number DE-SC0021117. E. F. and H.C. acknowledges the support from U.S. DOE BES Early Career Award KC0402010 under contract No. DE-AC05-00OR22725. This research used resources of the Advanced Light Source, which is a DOE Office of Science User Facility under contract No. DE-AC02-05CH11231.

\medskip

%

\end{document}